# X-ray attenuation of adipose breast tissue: In-vitro and in-vivo measurements using spectral imaging

Erik Fredenberg*[a], Klaus Erhard[b], Karl Berggren[a,c], David R. Dance[d,e], Kenneth C. Young [d,e], Björn Cederström[a], Henrik Johansson[a], Mats Lundqvist[a], Elin Moa[a], Hanno Homan[b], Paula Willsher[f], Fleur Kilburn-Toppin[f], Matthew Wallis[f]

[a] Philips Healthcare, Smidesvägen 5, 17141 Solna, Sweden;
[b] Philips Research, Röntgenstrasse 24, 22335 Hamburg, Germany;
[c] Department of Physics, Royal Institute of Technology (KTH), 10691 Stockholm, Sweden;
[d] NCCPM, Royal Surrey County Hospital, Guildford GU2 7XX, United Kingdom;
[e] Department of Physics, University of Surrey, Guildford GU2 7XH, United Kingdom;
[f] Cambridge Breast Unit and NIHR Cambridge Biomedical Research Centre, Addenbrookes Hospital, Hills Road, Cambridge CB2 0QQ, United Kingdom

**ABSTRACT**

The development of new x-ray imaging techniques often requires prior knowledge of tissue attenuation, but the sources of such information are sparse. We have measured the attenuation of adipose breast tissue using spectral imaging, in vitro and in vivo. For the in-vitro measurement, fixed samples of adipose breast tissue were imaged on a spectral mammography system, and the energy-dependent x-ray attenuation was measured in terms of equivalent thicknesses of aluminum and poly-methyl methacrylate (PMMA). For the in-vivo measurement, a similar procedure was applied on a number of spectral screening mammograms. The results of the two measurements agreed well and were consistent with published attenuation data and with measurements on tissue-equivalent material.

## 1. INTRODUCTION

Basic knowledge of tissue x-ray attenuation is essential for the development of new x-ray imaging techniques, as well as for estimating the performance of existing technologies. In the field of breast imaging, measurements have been reported of the attenuation of breast tumors[1–4] of fibroadenomas,[3] of cyst fluid,[5] and of adipose and glandular breast tissue.[1,4,2,3,6] There are, however, relatively large discrepancies between the different published measurements.

Knowledge of x-ray attenuation is particularly important for the development of new applications of spectral imaging, an emerging technology that measures the energy dependence of x-ray attenuation in the object. Unenhanced spectral imaging has been applied in mammography to improve the image signal-to-noise ratio,[7–9] improve lesion visibility,[10–13] discriminate between lesion types,[14,15] and measure breast density.[16] Common for the development of these applications is that accurate data of tissue attenuation is required, and the development is hampered by the discrepancies in published data.

We have previously developed a method to measure the energy-dependent x-ray attenuation of tissue samples using a clinical spectral-imaging mammography system.[5] This in-vitro method was applied in the present study to measure the attenuation of adipose tissue samples. Further, a similar method was developed and applied to measure the x-ray attenuation in adipose areas of screening mammograms, i.e. an in vivo measurement of tissue attenuation. These measurements of adipose tissue attenuation are a continuation of our efforts to characterize breast tissue and to resolve some of the discrepancies in the literature.

*erik.fredenberg@philips.com

## 2. METHODS

### 2.1 Measurement of x-ray attenuation with spectral imaging

For most natural body constituents at mammographic x-ray energies, it is fair to ignore absorption edges. X-ray attenuation is then made up of only two interaction effects, namely photoelectric absorption and scattering processes.[10,17,18] Accordingly, a combination of any two reference materials can approximately simulate the energy-dependent attenuation of a third material of certain thickness:

$$t_3 \times \mu_3(E) = t_1 \times \mu_1(E) + t_2 \times \mu_2(E). \tag{1}$$

If this relationship is assumed to hold exactly, measurements of the attenuation $t_3 \times \mu_3$ at two different energies on a tissue sample with known thickness $t_3$ are enough to uniquely determine the energy-dependent x-ray attenuation coefficient $\mu_3$ by solving for the thicknesses $t_1$ and $t_2$ of the two different reference materials with known attenuation coefficients $\mu_1$ and $\mu_2$. Measurements at more than two energies would yield an over-determined system of equations under the assumption of only two independent interaction processes, and would, in principle, be redundant. Equation (1) assumes that scattering processes can be treated as absorption, which is true only for x-ray detector geometries with efficient scatter rejection, such as multi-slit.[19]

### 2.2 Photon-counting spectral mammography system

The Philips MicroDose SI spectral mammography system (Philips Digital Mammography AB, Solna, Sweden) comprises an x-ray tube, a pre-collimator, and an image receptor, which is scanned across the object (Figure 1, left). The image receptor consists of photon-counting silicon strip detectors with corresponding slits in the pre-collimator (Figure 1, right). This multi-slit geometry rejects virtually all scattered radiation.

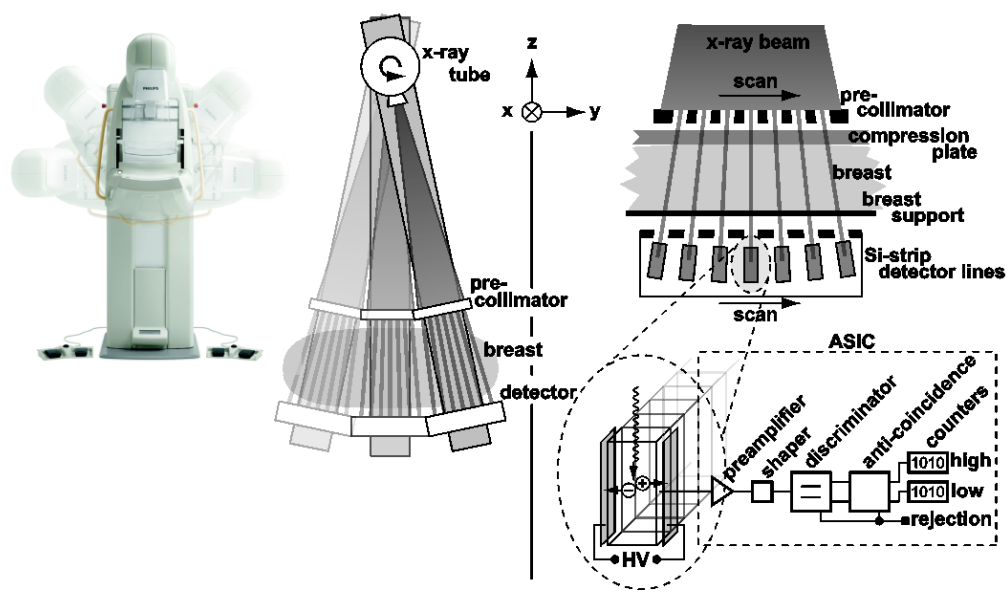

Figure 1: **Left:** Photograph and schematic of the Philips MicroDose SI Mammography system. **Right:** The spectral image receptor and electronics.

Photons that interact in the detector are converted to pulses with amplitude proportional to the photon energy.[20] Virtually all pulses below a few keV are generated by noise and are rejected by a low-energy threshold. A high-energy threshold sorts the detected pulses into two bins according to energy, and the system therefore enables the simultaneous measurement of the energy-dependent attenuation at two different effective energies.

### 2.3 Tissue characterization

#### 2.3.1 In-vitro measurements

Samples of fixed adipose breast tissue were obtained at the Cambridge Breast Unit, Cambridge, UK. Ethical approval was obtained to use samples from women from whom generic consent had been obtained prior to surgery. The tissue was sliced immediately post-surgery to facilitate even fixation with formalin. Six samples, 6-7 mm thick, were obtained from pathology the morning after surgery. The samples were placed in a hollow poly-methyl methacrylate (PMMA) cylinder with a threaded lid, which allowed gentle compression, henceforth referred to as the "sample holder". The sample thickness was measured using two methods: (1) A protractor on the lid allowed thickness measurement with a resolution of approximately 10 µm, and (2) the height of the cylinder was measured with a caliper, also with a resolution of 10 µm, from which the known bottom and lid thicknesses were subtracted. The mean of the two thickness measurements were used in the attenuation calculation. Figure 2 illustrates the measurement setup.

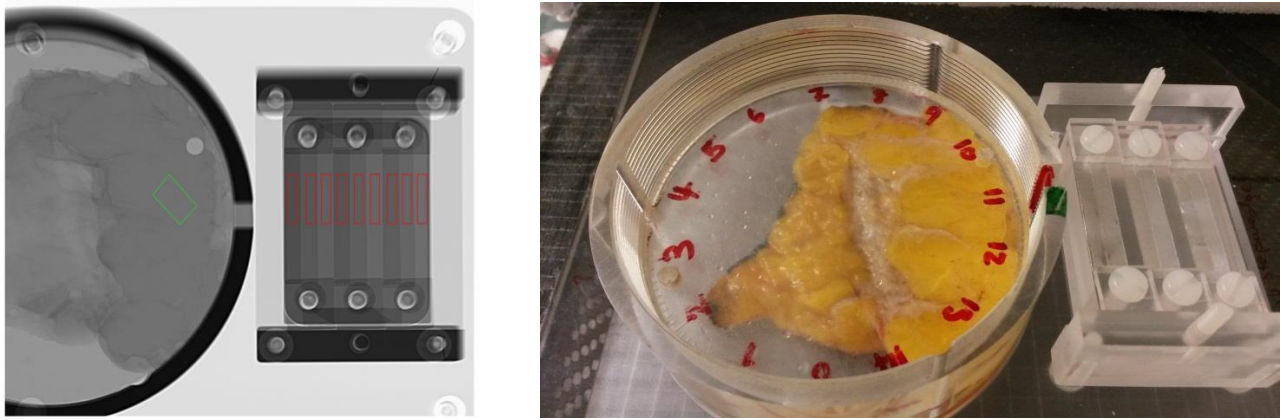

Figure 2: X-ray image (left) and photograph (right) of the sample holder and step wedge with an adipose sample in place. The sample ROI that was used for the evaluation is indicated in green and the step-wedge ROIs are indicated in red.

The samples with sample holder were imaged on a Philips MicroDose SI system. An aluminum (Al) and PMMA step wedge was present in each image adjacent to the sample to acquire a range of thickness/material combinations as a reference. X-ray attenuation was measured by mapping the high- and low-energy counts obtained from a region-of-interest (ROI) located on the sample against those obtained from ROIs on the Al-PMMA step wedge. Linear Delaunay triangulation in the log domain was used to find intermediate reference values. The in-vitro measurement procedure is described in more detail in Ref. 5.

In Ref. 5, the in-vitro measurement procedure was validated on five samples of breast cyst fluid using chemical analysis and elemental attenuation calculation as an independent method. The two methods agreed within the expected precision of the elemental attenuation calculation over the mammographic energy range.

Preliminary investigations revealed that the Al equivalence in the PMMA-Al description of adipose tissue attenuation is negative. To be able to measure the PMMA-Al equivalence, sheets of Al with well-defined thickness were added on top of the sample holder to reach a total positive Al thickness. The thickness of the Al sheets were subtracted from the value obtained from the step wedge.

#### 2.3.2 In-vivo measurements

For the in-vivo measurements, 94 anonymized spectral mammograms with low breast density were selected from a screening population that was examined with a Philips MicroDose SI system at Södersjukhuset, Stockholm, Sweden. Using the screening data for general research purposes was in accordance with local legislation.

The mammograms were automatically searched for fatty and homogeneous areas within the compressed region of the breast. ROIs were placed in these areas, which were assumed to contain nothing but adipose tissue. Similar to the in-vitro measurements, the spectral mammograms were calibrated so that high- and low-energy counts obtained from the ROIs could be mapped to equivalent thicknesses of Al and PMMA. The differences were, however, that the calibration was performed only once (there was no step wedge present in each image), the compression height was used to measure the thickness, and a fixed skin thickness of 3 mm was assumed. These simplifications can be expected to add a certain degree of systematic error to the measurement.

### 2.3.3 Comparison to reference data

For comparison, tissue attenuation in terms of Al and PMMA thicknesses was derived from four different publications.[1–3,6] The published values were converted to equivalent thicknesses of Al and PMMA using the methods described in Ref. 5. In addition, the attenuation of breast-tissue-equivalent phantom material (CIRS Inc., Norfolk, VA) was measured using the same procedure as the in-vitro measurements in Sec. 2.3.1.

## 3. RESULTS AND DISCUSSION

Figure 3 shows the equivalent thicknesses of Al and PMMA for 10 mm of normal breast tissue. The in-vitro and in-vivo measurements of adipose tissue are compared to values for adipose and glandular tissue derived from four different publications,[1–3,6] and derived from measurements and manufacturer specifications of tissue-equivalent material. The data are also listed in Table 1.

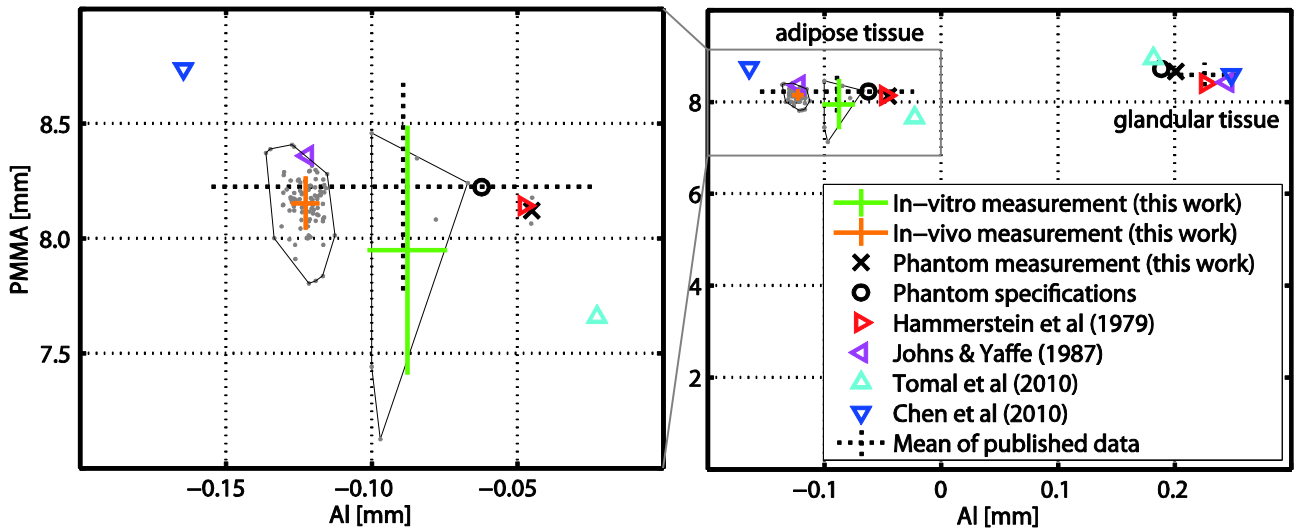

Figure 3: The equivalent thicknesses of aluminum (Al) and poly-methyl methacrylate (PMMA) for 10 mm of normal breast tissue. **Right:** Overview. **Left:** Close-up of the results for adipose tissue. Measurements on adipose tissue are shown for tissue samples (in vitro) and spectral screening mammograms (in vivo). Individual measurement points are shaded, colored error bars indicate the mean and one standard deviation of the spread, and the convex hulls of the measurement points are indicated by grey lines. Also shown are measurements on CIRS tissue-equivalent material (labeled phantom), equivalent thicknesses derived from published data on adipose and glandular tissue attenuation, and error bars for the mean and standard deviation of published data (not including phantom).

There was a significant difference between the in-vivo and in-vitro measurements ($p = 0.03$). The discrepancy can be attributed to a few systematic errors: tissue fixation in the in-vitro results, and contamination by skin, calibration error, and less accurate thickness estimation in the in-vivo results. Further, the accuracy of the in-vitro measurement could likely be improved by including a larger number of samples.

The results of both methods fell within the range of the published values. The in-vivo measurement was close to the values from Johns and Yaffe (differences: 0.6 μm Al, 210 μm PMMA),[1] whereas the in-vitro measurement was closer to the average of the four publications (differences: 1.5 μm Al, 280 μm PMMA).[1–3,6]

The spread of the published values is substantially larger for adipose tissue than for glandular tissue. This result is somewhat surprising as adipose tissue samples can be expected to be easier to extract surgically than glandular tissue and therefore contain less contamination by other tissue types.

The measurements on breast-equivalent material, adipose and glandular, agreed well with the material specifications, and with Hammerstein et al,[6] which the material was made to mimic. This result corroborates the in-vitro method in addition to the validation on cyst fluid that was presented in Ref. 5.

Table 1: Equivalent PMMA and Al thicknesses measured in 6 samples of adipose tissue (in vitro), measured in fatty areas of 94 spectral screening mammograms (in vivo), measured with the in-vitro method on CIRS tissue-equivalent material (labeled phantom), and calculated for adipose and glandular tissue using four sources of published data. The variability of the measurements is quantified as the coefficient of variation (the standard deviation divided by the mean, expressed as a percentage) between samples or mammograms. The variability of the published data is quantified as the coefficient of variation between the different sources.

|  | equivalent thickness of PMMA | | equivalent thickness of Al | |
|---|---|---|---|---|
|  | mean | variability | mean | variability |
| measured adipose in vitro | 7.95 mm | 6.8% | -0.088 mm | 15% |
| measured adipose in vivo | 8.15 mm | 1.4% | -0.123 mm | 3.9% |
| measured adipose phantom | 8.12 mm | - | -0.045 mm | - |
| published adipose | 8.22 mm | 5.5% | -0.089 mm | 73% |
| measured glandular phantom | 8.66 mm | - | 0.201 mm | - |
| published glandular | 8.59 mm | 2.9% | 0.226 mm | 14% |

## 4. CONCLUSIONS

Published data on normal tissue attenuation cover a relatively wide range, in particular for adipose tissue. Our study provides new measurements of the attenuation of adipose tissue, which were performed under conditions that are as similar as possible to the screening environment, and which may serve to reduce some of the systematic uncertainty in the literature. Further, the in-vivo method provides a new non-invasive way of finding values of tissue attenuation, which opens up for including large data sets and may reduce statistical errors. More accurate data on breast-tissue attenuation would ultimately improve the accuracy of advanced techniques in mammography, such as measurement of breast-density[16] and discrimination between lesion types.[14,15]

There was a significant difference between the in-vitro and in-vivo results, which could, however, be attributed to a few systematic errors. The results of both methods fell within the range of values from the literature. The in-vitro measurements of tissue-equivalent material were consistent with published values. Potential future improvements include improved thickness estimates for the in-vivo measurement, and using a larger number of samples of fresh (rather than fixed) tissue for the in-vitro measurement.

## ACKNOWLEDGEMENTS

This work was supported by Cancer Research UK as part of the OPTIMAM2 project (A17321).

## REFERENCES

[1] Johns, P. C., Yaffe, M. J., "X-ray characterisation of normal and neoplastic breast tissues," Phys. Med. Biol. **32**(6), 675-695 (1987).
[2] Chen, R. C., Longo, R., Rigon, L., Zanconati, F., De Pellegrin, A., Arfelli, F., Dreossi, D., Menk, R. H., Vallazza, E., Xiao, T. Q., Castelli, E., "Measurement of the linear attenuation coefficients of breast tissues by synchrotron radiation computed tomography," Phys. Med. Biol. **55**(17), 4993-5005 (2010).
[3] Tomal, A., Mazarro, I., Kakuno, E. M., Poletti, M. E., "Experimental determination of linear attenuation coefficient of normal, benign and malignant breast tissues," Radiat. Meas. **45**(9), 1055-1059 (2010).
[4] Carroll, F. E., Waters, J. W., Andrews, W. W., Price, R. R., Pickens, D. R., Willcott, R., Tompkins, P., Roos, C., Page, D., Reed, G., Ueda, A., Bain, R., Wang, P., Bassinger, M., "Attenuation of Monochromatic X-Rays by Normal and Abnormal Breast Tissues," Investigative Radiology **29**(3), 266-272 (1994).

[5] Fredenberg, E., Dance, D. R., Willsher, P., Moa, E., von Tiedemann, M., Young, K. C.., Wallis, M. G., "Measurement of breast-tissue x-ray attenuation by spectral mammography: first results on cyst fluid," Phys. Med. Biol. **58**(24), 8609–8620 (2013).

[6] Hammerstein, G. R., Miller, D. W., White, D. R., Masterson, M. E., Woodard, H. Q., Laughlin, J. S., "Absorbed Radiation-Dose in Mammography," Radiology **130**(2), 485-491 (1979).

[7] Tapiovaara, M. J., Wagner, R. F., "SNR and DQE analysis of broad spectrum x-ray imaging," Phys. Med. Biol. **30**, 519-529 (1985).

[8] Cahn, R. N., Cederström, B., Danielsson, M., Hall, A., Lundqvist, M., Nygren, D., "Detective quantum efficiency dependence on x-ray energy weighting in mammography," Med. Phys. **26**(12), 2680-2683 (1999).

[9] Berglund, J., Johansson, H., Lundqvist, M., Cederström, B.., Fredenberg, E., "Energy weighting improves dose efficiency in clinical practice: implementation on a spectral photon-counting mammography system," J. Med. Imag. **1**(3), 031003 (2014).

[10] Johns, P. C., Yaffe, M. J., "Theoretical optimization of dual-energy x-ray imaging with application to mammography," Med. Phys. **12**, 289-296 (1985).

[11] Fredenberg, E., Åslund, M., Cederström, B., Lundqvist, M., Danielsson, M. E., "Observer model optimization of a spectral mammography system," Proc of SPIE Medical Imaging 2010: Physics of Medical Imaging, (2010).

[12] Kappadath, S. C., Shaw, C. C., "Quantitative evaluation of dual-energy digital mammography for calcification imaging," Phys. Med. Biol. **53**(19), 5421-5443 (2008).

[13] Taibi, A., Fabbri, S., Baldelli, P., di Maggio, C., Gennaro, G., Marziani, M., Tuffanelli, A., Gambaccini, M., "Dual-energy imaging in full-field digital mammography: a phantom study," Phys. Med. Biol. **48**, 1945-1956 (2003).

[14] Norell, B., Fredenberg, E., Leifland, K., Lundqvist, M., Cederström, B., "Lesion characterization using spectral mammography," Proc of SPIE Medical Imaging 2012: Physics of Medical Imaging, (2012).

[15] Erhard, K., Fredenberg, E., Homann, H., Roessl, E., "Spectral lesion characterization on a photon-counting mammography system," Proc of SPIE Medical Imaging 2014: Physics of Medical Imaging, (2014).

[16] Ding, H., Molloi, S., "Quantification of breast density with spectral mammography based on a scanned multi-slit photon-counting detector: a feasibility study," Phys. Med. Biol. **57**(15), 4719-4738 (2012).

[17] Alvarez, R. E., Macovski, A., "Energy-Selective Reconstructions in X-Ray Computerized Tomography," Phys. Med. Biol. **21**(5), 733-744 (1976).

[18] Lehmann, L. A., Alvarez, R. E., Macovski, A., Brody, W. R., Pelc, N. J., Riederer, S. J., Hall, A. L., "Generalized image combinations in dual KVP digital radiography," Med. Phys. **8**(5), 659-667 (1981).

[19] Åslund, M., Cederström, B., Lundqvist, M., Danielsson, M., "Scatter rejection in multi-slit digital mammography," Med. Phys. **33**, 933-940 (2006).

[20] Fredenberg, E., Lundqvist, M., Cederström, B., Åslund, M., Danielsson, M., "Energy resolution of a photon-counting silicon strip detector," Nucl. Instr. and Meth. A **613**(1), 156-162 (2010).